\documentclass[english]{iopart}
\usepackage[T1]{fontenc}
\usepackage[latin9]{inputenc}
\usepackage{graphicx}
\usepackage{amssymb}

\newcommand{\noun}[1]{\textsc{#1}}

\usepackage{iopams}
\usepackage{setstack}

\usepackage{dsfont}

\usepackage{babel}

\usepackage{babel}

\begin{document}

\title{Spatial distribution of Cherenkov radiation in periodic dielectric
media}

\author{Christian Kremers and Dmitry N. Chigrin}

\address{Theoretical Nano-Photonics, Institute of High-Frequency and Communication
Technology, Faculty of Electrical, Information and Media Engineering,
University of Wuppertal, Rainer-Gruenter-Str. 21, D-42119 Wuppertal,
Germany}

\ead{kremers@uni-wuppertal.de}
\begin{abstract}
The nontrivial dispersion relation of a periodic medium affects both
the spectral and the spatial distribution of Cherenkov radiation.
We present a theory of the spatial distribution of Cherenkov radiation
in the far-field zone inside arbitrary three- and two-dimensional
dielectric media. Simple analytical expressions for the far-field
are obtained in terms of the Bloch mode expansion. Numerical examples
of the Cherenkov radiation in a two-dimensional photonic crystal is
presented. The developed analytical theory demonstrates good agreement
with numerically rigorous finite-difference time-domain calculations. 
\end{abstract}
\maketitle

\section{Introduction}

In recent years light-matter interaction in inhomogeneous media, structured
on the sub-wavelength scale, has attracted a considerable attention.
The interplay between interference and propagation can result in a
nontrivial dispersion relation in such a medium. For example, in a
periodic dielectric medium \cite{jmw95,sak01} both dispersion and
diffraction of electromagnetic waves are substantially modified leading
to novel optical phenomena, including the emission dynamics modification
\cite{byk72,yab87,jw90}, ultra-refraction \cite{rus86,zen87,kos98,kos99}
and photon focusing \cite{ep96,cst01,chi04} effects. A nontrivial
dispersion relation can also substantially modify the Cherenkov radiation
\cite{che34,jel58,fer40}. For example an electron moving in a homogeneous
medium with dispersion can emit at any velocity \cite{fer40} and
the spatial distribution of the emitted radiation demonstrates intensity
oscillations behind the Cherenkov cone \cite{akm99,carb01,acrb03}.

Periodic media modify both spectral \cite{Luo03,kremers09} and spatial
\cite{Luo03} distribution of the Cherenkov radiation. At the moment,
several studies on the modification of the Cherenkov and Smith-Purcell
radiation are available. The Smith-Purcell radiation has been studied
in Refs.~\cite{abajo03b,oo04a,oo04b,yam04,oo06}. The Cherenkov radiation
has been used to map a band structure of a two-dimentional (2D) photonic
crystal in Refs.~\cite{abajo03,abajo03a}. We have recently presented
a general theory of spectral power modification in both three-dimensional
(3D) and 2D periodic media \cite{kremers09}. Spatial modifications
of the Cherenkov radiation have been numerically analyzed in Ref.~\cite{Luo03}
using the finite-difference time-domain (FDTD) method.

The main purpose of the present work is to develop an analytical theory
of the Cherenkov radiation in far-field zone in order to provide a
simple semi-analytical tool to study its spatial distribution pecularities
in general 3D and 2D periodic dielectric media. To achieve this goal
we first derive an analytical expression for the electric far-field
in terms of the Bloch mode expansion. Secondly we restrict the full
$k$-space integration in the Bloch mode expansion of the field to
a relatively simple surface (contour) integral in the first Brillouin
zone of the 3D (2D) periodic medium. Thirdly, we restrict this integral
further to a simple sum over a small number of Bloch eigenmodes. These
derived formulas allow us to identify the main contribution to the
spatial pecularities of the Cherenkov radiation.

The paper is organized as follows: In Section \ref{sec:Radiated-field}
the general solution of Maxwell's equations is summarized. In Section
\ref{sec:far_field_approximation} an analytical expression for the
Cherenkov far-field is derived both for 3D and 2D periodic dielectric
media. In Section \ref{sec:Numerical-example} we apply the developed
theory to calculate the Cherenkov far-field in the particular case
of a 2D photonic crystal. Predictions of the analytical theory are
substantiated by numerically rigorous FDTD calculations. Section \ref{sec:Conclusions}
concludes the paper.

\section{General solution\label{sec:Radiated-field}}

\begin{figure}
\begin{centering}
\includegraphics[width=0.5\textwidth]{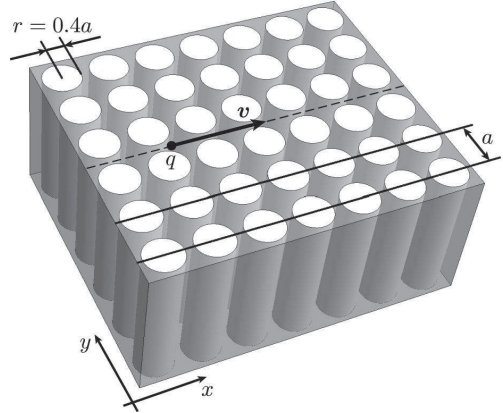} 
\par\end{centering}

\caption{A sketch of a periodic medium and charge trajectory. Parameters of
the infinite 2D photonic crystal used in the numerical example (Sec.
\ref{sec:Numerical-example}) are shown.\label{fig:sketch}}

\end{figure}

We consider a point (line) charge $q$ uniformly moving with a velocity
$\bi{v}$ in an infinite periodic 3D (2D) dielectric medium (Fig.~\ref{fig:sketch}).
The medium is described by a periodic dielectric function $\varepsilon\left(\bi{r}\right)=\varepsilon\left(\bi{\bi{r}}+\bi{R}\right)$
with $\bi{R}=\sum_{i}l_{i}\bi{a}_{i}$ being a vector of the direct
Bravais lattice. The $l_{i}$ are integers and the $\bi{a}_{i}$ are
basis vectors of the lattice. We assume also that the medium is linear,
nonmagnetic and loss less. The relevant part of Maxwell's equations
read in SI units: \begin{eqnarray}
\nabla\times\bi{E}\left(\bi{r},t\right) & = & -\mu_{0}\frac{\partial}{\partial t}\bi{H}\left(\bi{r},t\right),\label{eq:maxwell_rotE}\\
\nabla\times\bi{H}\left(\bi{r},t\right) & = & \varepsilon_{0}\varepsilon\left(\bi{r}\right)\frac{\partial}{\partial t}\bi{E}\left(\bi{r},t\right)+\bi{J}\left(\bi{r},t\right),\label{eq:maxwell_rotH}\end{eqnarray}
 where $\bi{E}(\bi{r},t)$ is the electric field, $\bi{H}(\bi{r},t)$
is the magnetic field and $\bi{J}(\bi{r},t)$ is the current density.
In the presence of an arbitrary current density switched on adiabatically
$\bi{J}(\bi{r},t\rightarrow-\infty)=0$, a general solution of Maxwell's
equations (\ref{eq:maxwell_rotE}-\ref{eq:maxwell_rotH}) for the
electric field at point $\bi{r}$ and time $t$ can be given in terms
of the Bloch eigenmode expansion \cite{sak01,dow92} \begin{eqnarray}
\fl\bi{E}(\bi{r},t)=-\frac{1}{(2\pi)^{d}\varepsilon_{0}}\sum_{n}\int_{{\rm {BZ}}}\rmd^{d}k\int\rmd^{d}r'\int_{-\infty}^{t}\rmd t'\,\left\{ \cos\left[\omega_{\bi{k}n}^{(T)}\left(t-t'\right)\right]\right.\nonumber \\
\cdot\bi{E}_{\bi{k}n}^{(T)}\left(\bi{r}\right)\otimes\bi{E}_{\bi{k}n}^{(T)\star}\left(\bi{r}'\right)+\left.\bi{E}_{\bi{k}n}^{(L)}\left(\bi{r}\right)\otimes\bi{E}_{\bi{k}n}^{(L)\star}\left(\bi{r}'\right)\right\} \cdot\bi{J}(\bi{r}',t').\label{eq:E(r,t)_general}\end{eqnarray}
 $d=2,\,3$ is the dimensionality of the periodic lattice. The asterisk
($\star$) and $\otimes$ denote the complex conjugate and the outer
tensor product, respectively. $\bi{E}_{\bi{k}n}^{(T)}\left(\bi{r}\right)$
and $\bi{E}_{\bi{k}n}^{(L)}\left(\bi{r}\right)$ are generalized transverse
and longitudinal Bloch eigenmodes \cite{sak01,dow92} characterized
by the band index $n$, the wave vector $\bi{k}$ and the eigenfrequencies
$\omega_{\bi{k}n}^{(T)}$ and $\omega_{\bi{k}n}^{(L)}$, where $\omega_{\bi{k}n}^{(L)}=0$.
The $k$-space integration is performed over the first Brillouin zone
(${\rm {BZ}}$) of the periodic lattice and the summation is carried
out over the different photonic bands. Bloch eigenmodes are solutions
of the homogeneous wave equation and satisfy periodic boundary conditions.
These eigenmodes are normalized \begin{equation}
\int\rmd^{d}r\,\varepsilon\left(\bi{r}\right)\bi{E}_{\bi{k}n}^{(\alpha)\star}\left(\bi{r}\right)\cdot\bi{E}_{\bi{k}'n'}^{(\beta)}\left(\bi{r}\right)=\left(2\pi\right)^{d}\delta_{\alpha\beta}\delta_{nn'}\delta^{(d)}\left(\bi{k}-\bi{k}'\right)\label{eq:normalisation}\end{equation}
 and satisfy the completeness relation \begin{eqnarray}
\sum_{n\alpha}\int_{{\rm {BZ}}}\rmd^{d}k\,\sqrt{\varepsilon\left(\bi{r}\right)\varepsilon\left(\bi{r}'\right)}\bi{E}_{\bi{k}n}^{(\alpha)}\left(\bi{r}\right)\otimes\bi{E}_{\bi{k}n}^{(\alpha)\star}\left(\bi{r}'\right)\nonumber \\
=\left(2\pi\right)^{d}\mathds{1}\delta^{(d)}\left(\bi{r}-\bi{r}'\right),\label{eq:completeness}\end{eqnarray}
 where $\alpha,\,\beta=T$ or $L$, and $\mathds{1}$ is the unit
tensor.

For a point (line) charge $q$ moving with the velocity $\bi{v}$,
the current density $\bi{J}(\bi{r},t)$ can be expressed as \begin{equation}
\bi{J}\left(\bi{r},t\right)=q\bi{v}\delta^{(d)}\left(\bi{r}-\bi{v}t\right).\label{eq:current_time}\end{equation}
 Using this current density and performing the space integration in
the Bloch eigenmode expansion (\ref{eq:E(r,t)_general}) the following
general solution can be obtained\begin{eqnarray}
\fl\bi{E}(\bi{r},t)=-\frac{q\left|\bi{v}\right|}{(2\pi)^{d}\varepsilon_{0}}\sum_{n}\int_{{\rm {BZ}}}\rmd^{d}k\int_{-\infty}^{t}\rmd t'\,\cos\left[\omega_{\bi{k}n}^{(T)}\left(t-t'\right)\right]\nonumber \\
\cdot\bi{E}_{\bi{k}n}^{(T)}\left(\bi{r}\right)\left(\bi{E}_{\bi{k}n}^{(T)\star}\left(\bi{v}t'\right)\cdot\hat{\bi{v}}\right)\label{eq:E(r,t)_transverse_exact}\end{eqnarray}
 with $\hat{\bi{v}}=\bi{v}/\left|\bi{v}\right|$ being the unit vector
in the direction of the charge velocity. Here the fact that the longitudinal
modes do not contribute to the radiated field \cite{kremers09} has
been taken into account. In what follows the upper index $(T)$ denoting
transverse eigenmodes and eigenfrequencies will be dropped.

\section{Cherenkov radiation in the far-field zone\label{sec:far_field_approximation}}

Expressing the cosine function in Eq.~(\ref{eq:E(r,t)_transverse_exact})
as a sum of two complex exponential functions and using Bloch's theorem
$\bi{E}_{\bi{k}n}(\bi{r})=\bi{e}_{\bi{k}n}\left(\bi{r}\right)\exp\left(\rmi\bi{k}\cdot\bi{r}\right)$,
where $\bi{e}_{\bi{k}n}\left(\bi{r}\right)$ is a lattice periodic
function \cite{sak01}, a general solution for the Cherenkov field
can be rewritten in the form \begin{equation}
\bi{E}(\bi{r},t)=-\frac{q\left|\bi{v}\right|}{2(2\pi)^{d}\varepsilon_{0}}\sum_{n}\left(I_{+}+I_{-}\right)\label{eq:E(r,t)_bloch_exact_without_fouriercoeff}\end{equation}
 where $I_{\pm}$ is defined as \begin{equation}
I_{\pm}=\int_{{\rm {BZ}}}\rmd^{d}k\int_{-\infty}^{t}\rmd t'\,\bi{e}_{\bi{k}n}\left(\bi{r}\right)\left(\bi{e}_{\bi{k}n}^{\star}\left(\bi{v}t'\right)\cdot\hat{\bi{v}}\right)\mathcal{E}_{\pm}(\bi{k},t')\label{eq:E(r,t)_bloch_exact_without_fouriercoeff_2}\end{equation}
 with \begin{equation}
\mathcal{E}_{\pm}(\bi{k},t')=\exp\left\{ \rmi\left[\bi{k}\cdot\left(\bi{r}-\bi{v}t'\right)\pm\omega_{\bi{k}n}\left(t-t'\right)\right]\right\} .\end{equation}
 Further taking into account the symmetries of the Bloch eigenmodes,
$\bi{e}_{-\bi{k}n}=\bi{e}_{\bi{k}n}^{\star}$ and $\omega_{-\bi{k}n}=\omega_{\bi{k}n}$
\cite{sak01}, the following relation for the integrals $I_{\pm}$
(\ref{eq:E(r,t)_bloch_exact_without_fouriercoeff_2}) holds \begin{eqnarray}
I_{-}^{\star} & =\int_{{\rm {BZ}}}\rmd^{d}k\int_{-\infty}^{t}\rmd t'\,\bi{e}_{-\bi{k}n}\left(\bi{r}\right)\left(\bi{e}_{-\bi{k}n}^{\star}\left(\bi{v}t'\right)\cdot\hat{\bi{v}}\right)\mathcal{E}_{+}(-\bi{k},t')\nonumber \\
 & =I_{+}.\end{eqnarray}
 Therefore the radiated field (\ref{eq:E(r,t)_bloch_exact_without_fouriercoeff})
can be exclusively expressed in terms of the real part of the integral
$I_{-}$ \begin{equation}
\bi{E}(\bi{r},t)=-\frac{q\left|\bi{v}\right|}{(2\pi)^{d}\varepsilon_{0}}\sum_{n}{\rm {Re}}\left(I_{-}\right).\label{eq:E_final_I_minus}\end{equation}

To analyze further the Cherenkov field (\ref{eq:E_final_I_minus})
we limit ourselves to the charge trajectories which do not cut dielectric
interfaces of the periodic medium. In this case {}``bremsstrahlung''
radiation can be neglected and the trajectories themselves are necessarily
oriented rationally with respect to the periodic lattice. In this
case $\bi{e}_{\bi{k}n}^{\star}\left(\bi{v}t'\right)\cdot\hat{\bi{v}}$
in (\ref{eq:E(r,t)_bloch_exact_without_fouriercoeff_2}) is a one-dimensional
(1D) periodic function with a period $\mathfrak{a}$ defined by the
orientation of the charge trajectory therefore one can Fourier expand
it as follows \begin{equation}
\bi{e}_{\bi{k}n}^{\star}(\bi{v}t')\cdot\hat{\bi{v}}=\sum_{m=-\infty}^{\infty}c_{nm}(\bi{k})\exp\left(\rmi\frac{2\pi|\bi{v}|}{\mathfrak{a}}mt'\right)\label{eq:fourier_sum_of_blochmode}\end{equation}
 with Fourier coefficients \begin{equation}
c_{nm}(\bi{k})=\frac{1}{\mathfrak{a}}\int_{0}^{\mathfrak{a}}d\xi\,\left(\bi{e}_{\bi{k}n}^{\star}(\xi\hat{\bi{v}})\cdot\hat{\bi{v}}\right)\exp\left(-\rmi\frac{2\pi}{\mathfrak{a}}m\xi\right).\label{eq:def_of_c_m}\end{equation}
 This allows us to rewrite (\ref{eq:E(r,t)_bloch_exact_without_fouriercoeff_2})
in the form \begin{equation}
I_{-}=\sum_{m=-\infty}^{\infty}\int_{{\rm {BZ}}}\rmd^{d}k\int_{-\infty}^{t}\rmd t'\,\bi{e}_{\bi{k}n}\left(\bi{r}\right)c_{nm}(\bi{k})\tilde{\mathcal{E}}_{nm}(\bi{k},t')\label{eq:def_I(pm)}\end{equation}
 with function $ $$\tilde{\mathcal{E}}_{nm}(\bi{k},t')$ defined
by \begin{eqnarray}
\tilde{\mathcal{E}}_{nm}(\bi{k},t')=\exp\left[\rmi\left(\bi{k}\cdot\left(\bi{r}-\bi{v}t'\right)-\omega_{\bi{k}n}\left(t-t'\right)+\frac{2\pi|\bi{v}|}{\mathfrak{a}}mt'\right)\right]\nonumber \\
=\exp\left[\rmi f_{nm}(\bi{k},t')\right].\label{eq:def_Etilde_and_fnm}\end{eqnarray}

We are interested in the field far away from the trajectory of the
charge. In the far-field zone the following relation $|\bi{r}-\bi{v}t'|\gg\lambda$
holds for all moments of time $t'\leq t$. If this condition is fulfilled
a small variation of wave vector $\bi k$ results in rapid oscillations
of the exponential function $\tilde{\mathcal{E}}_{nm}(\bi{k},t')$.
Taking into account that the function $\bi{e}_{\bi{k}n}\left(\bi{r}\right)c_{nm}(\bi{k})$
is a slow function of the wave vector, the main contribution to the
integral $I_{-}$ in the far-field zone comes from the neighborhood
of $k$-points where the variation of the phase $f_{nm}(\bi k,t')$
is minimal. Such stationary $k$-points are defined by the relation
\begin{equation}
\nabla_{\bi{k}}f_{nm}(\bi{k},t')=0,\label{eq:def_stat_points}\end{equation}
 which explicitly reads as\begin{equation}
\bi{v}_{{\rm {g}}}(\bi{k})=\nabla_{\bi{k}}\omega_{n}(\bi{k})=\frac{\bi{r}-\bi{v}t'}{t-t'}.\label{eq:stationary_2}\end{equation}
 Where $\bi{v}_{{\rm {g}}}(\bi{k})$$ $ is the group velocity of
the Bloch eigenmode. Relation (\ref{eq:stationary_2}) can be written
in the equivalent form\begin{equation}
\left(\bi{v}_{{\rm {g}}}(\bi{k})-\bi{v}\right)\cdot\left(\bi{r}-\bi{v}t\right)=\frac{\left|\bi{r}-\bi{v}t\right|^{2}}{t-t'},\label{eq:stationary_3}\end{equation}
 with its right hand side being positive for all moments of time $t'<t$.
Taking that into account, the integration in (\ref{eq:def_I(pm)})
can be restricted to the part of the Brillouin zone, ${\rm BZ}_{1}$,
containing all wave vectors whose group velocities fulfill the relation\begin{equation}
\left(\bi{v}_{{\rm {g}}}(\bi{k})-\bi{v}\right)\cdot\left(\bi{r}-\bi{v}t\right)>0.\label{eq:stationary_3_ie}\end{equation}
 Further, noting that in ${\rm BZ}_{1}$ there does not exist any
stationary points for $t'>t$, the $t'$-integration in (\ref{eq:def_I(pm)})
can be extended to the whole real axis without severe error. With
a good accuracy the integral $I_{-}$ can be approximated by\begin{equation}
I_{-}\approx\sum_{m=-\infty}^{\infty}\int_{{\rm {BZ}}_{1}}\rmd^{d}k\int_{-\infty}^{\infty}\rmd t'\,\bi{e}_{\bi{k}n}\left(\bi{r}\right)c_{nm}(\bi{k})\tilde{\mathcal{E}}_{nm}(\bi{k},t').\end{equation}
 Using the integral expression of the Dirac delta function, the $t'$-integration
can be easily performed resulting in\begin{eqnarray}
I_{-} & \approx & 2\pi\sum_{m=-\infty}^{\infty}\int_{{\rm {BZ}}_{1}}\rmd^{d}k\,\bi{e}_{\bi{k}n}(\bi{r})c_{nm}(\bi{k})\exp\left[\rmi\left(\bi{k}\cdot\bi{r}-\omega_{\bi{k}n}t\right)\right]\nonumber \\
 &  & \quad\cdot\delta\left(\omega_{\bi{k}n}-\bi{k}\cdot\bi{v}+\frac{2\pi\left|\bi{v}\right|}{\mathfrak{a}}m\right).\label{eq:I_minus_k-t}\end{eqnarray}
 Furthermore, using the integral relation\[
\int_{V}\rmd^{d}r\, f(\bi{r})\delta\left(g\left(\bi{r}\right)\right)=\int_{\partial V}\rmd^{d-1}r\,\frac{f(\bi{r})}{\left|\nabla g(\bi{r})\right|},\]
 where $\partial V$ is a surface defined by the equation $g\left(\bi{r}\right)=0$$ $,
the $d$-dimensional integral in (\ref{eq:I_minus_k-t}) can be reduced
to the $(d-1)$-dimensional integral \begin{equation}
I_{-}\approx2\pi\sum_{m}\int_{\mathcal{C}_{m}}\rmd^{d-1}k\,\frac{\bi{e}_{\bi{k}n}\left(\bi{r}\right)c_{nm}(\bi{k})}{\left|\bi{v}_{{\rm {g}}}(\bi{k})-\bi{v}\right|}\exp\left[\rmi\left(\bi{k}\cdot\bi{r}-\omega_{\bi{k}n}t\right)\right],\label{eq:I_minus_final}\end{equation}
 which finally gives the following expression for the Cherenkov far-field\begin{eqnarray}
\bi{E}(\bi{r},t) & = & -\frac{q\left|\bi{v}\right|}{(2\pi)^{d-1}\varepsilon_{0}}\sum_{n,m}{\rm {Re}}\Biggl\{\int_{\mathcal{C}_{m}}\rmd^{d-1}k\,\frac{\bi{e}_{\bi{k}n}\left(\bi{r}\right)c_{nm}(\bi{k})}{\left|\bi{v}_{{\rm {g}}}(\bi{k})-\bi{v}\right|}\nonumber \\
 &  & \Biggl.\left.\cdot\exp\left[\rmi\left(\bi{k}\cdot\bi{r}-\omega_{\bi{k}n}t\right)\right]\right)\Biggr\}.\label{eq:E_final}\end{eqnarray}
 The integration is performed over the part of the Brillouin zone,
${\rm BZ}_{1}$, defined by (\ref{eq:stationary_3_ie}) and the integration
surface (3D case) or contour (2D case) $\mathcal{C}_{m}$ is defined
by the generalized Cherenkov condition \cite{kremers09}\begin{equation}
\omega_{\bi{k}n}=\bi{k}\cdot\bi{v}-\frac{2\pi}{\mathfrak{a}}m\left|\bi{v}\right|,\label{eq:Cherenkov_condition}\end{equation}
 where $m$ is an integer. In what follows we will refer to the integration
surface (contour) (\ref{eq:Cherenkov_condition}) as Cherenkov surface
(contour).

\begin{figure}
\begin{centering}
\includegraphics[width=0.5\textwidth]{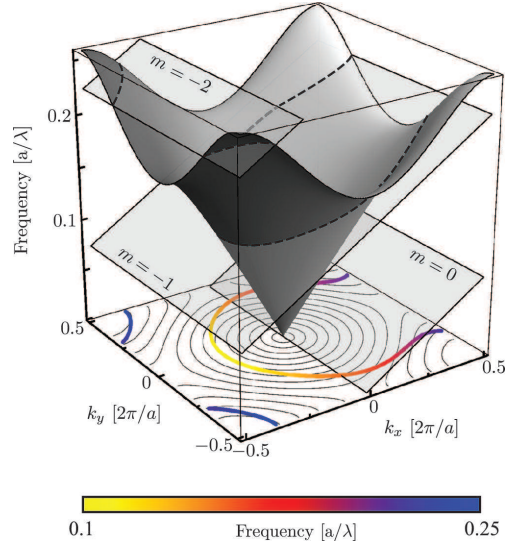} 
\par\end{centering}

\caption{Diagram to illustrate the generalized Cherenkov condition. The band
structure of the 2D photonic crystal (Fig: \ref{fig:band_structure})
for TE (transverse electric) polarization as well as the set of planes
for $m=0$, $m=-1$ and $m=-2$ and the charge velocity $\bi{v}=0.15c\,\hat{\bi{x}}$
are shown. The intersections of the band structure with the planes
define the integration contour $\mathcal{C}_{m}$ (dashed lines).
A color coded projection of the contour on the first Brillouin zone
is also shown.\label{fig:band_structure}}

\end{figure}

Equation (\ref{eq:E_final}) is the main result of the present section.
We have demonstrated that a restriction of the integration range from
the whole Brillouin zone to the solutions of the generalized Cherenkov
condition (\ref{eq:Cherenkov_condition}) in ${\rm BZ}_{1}$ (\ref{eq:stationary_3_ie})
is possible. The generalized Cherenkov condition (\ref{eq:Cherenkov_condition})
chooses all Bloch modes contributing to the Cherenkov radiation. A
graphical illustration of the Cherenkov condition (\ref{eq:Cherenkov_condition})
is presented in figure~\ref{fig:band_structure}. The band structure
of an infinite square lattice photonic crystal is shown. The dielectric
constant of the background medium is $\varepsilon=12$ and the radius
of the air holes is $r=0.4a$, where $a$ is the lattice constant
(Fig.~\ref{fig:sketch}). The manifold of the Cherenkov wave vectors
(the integration contour $\mathcal{C}_{m}$) is given by the intersection
of the band structure, $\omega_{\bi{k}n}$, with the set of planes
$f(\bi{k}_{\parallel})=\left|\bi{v}\right|\bi{k}_{\parallel}-\left|\bi{v}\right|\frac{2\pi}{\mathfrak{a}}m$
for different $m$. Here $\bi{k}_{\parallel}$ is the component of
the wave vector parallel to the charge velocity. The slope of the
planes is defined by the charge velocity, being $\bi{v}=0.15c\,\hat{\bi{x}}$
in this example.

\subsection{2D periodic media}

\begin{figure}
\begin{centering}
\includegraphics[width=0.5\textwidth]{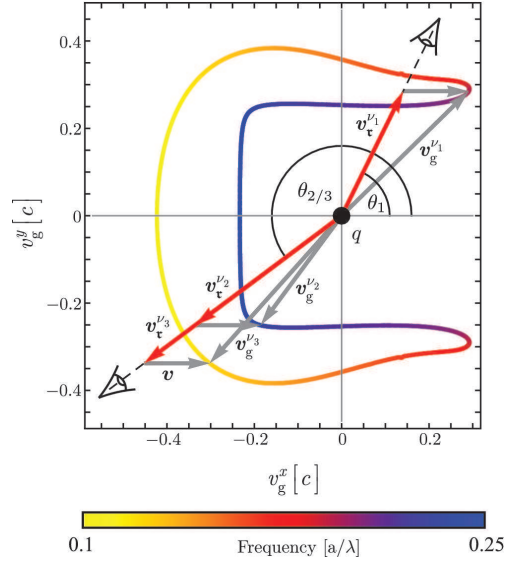} 
\par\end{centering}

\caption{Group velocity contour corresponding to the integration contour $\mathcal{C}_{m}$.
The main contribution to the Cherenkov radiation in the direction
$\theta_{1}$ ($\theta_{2/3}$) comes from the Bloch mode(s) with
group velocity (velocities) $\bi v_{{\rm g}}^{\nu_{1}}$ ($\bi v_{{\rm g}}^{\nu_{2}}$
and $\bi v_{{\rm g}}^{\nu_{3}}$). The same color coding as in Fig.~\ref{fig:band_structure}
is used.\label{fig:group_velocity}}

\end{figure}

To further simplify the integral (\ref{eq:I_minus_final}) we can
parameterize the contour $\mathcal{C}_{m}$ by its arc length $s$.
Then the contour integral (\ref{eq:I_minus_final}) with $d=2$ can
be transformed into an integral over $s$\begin{eqnarray}
I_{-} & \approx & 2\pi\sum_{m}\int\rmd s\,\frac{\bi{e}_{n}\left(\bi{k}(s),\bi{r}\right)c_{nm}(\bi{k}(s))}{\left|\bi{v}_{{\rm {g}}}(\bi{k}(s))-\bi{v}\right|}\nonumber \\
 &  & \cdot\exp\left\{ \rmi\left[\bi{k}(s)\cdot\bi{r}-\omega_{n}(\bi{k}(s))t\right]\right\} .\label{eq:I_minus_para_1}\end{eqnarray}
 The main contribution to the integral comes from the \emph{k}-points
$\bi{k}(s^{\nu})$ in whose neighborhood the phase $h\left(s\right)=\bi{k}(s)\cdot\bi{r}-\omega_{n}(\bi{k}(s))t$
is stationary with respect to the variation of $s$\begin{eqnarray}
\left.\frac{\partial}{\partial s}\left[\bi{k}(s)\cdot\bi{r}-\omega_{n}(\bi{k}(s))t\right]\right|_{s=s^{\nu}} & = & 0\nonumber \\
\left(\bi{r}-\bi{v}t\right)\cdot\frac{\partial\bi{k}^{\nu}}{\partial s} & = & 0.\label{eq:stat_4}\end{eqnarray}
 The second equality in (\ref{eq:stat_4}) holds as the derivative
of the Cherenkov condition (\ref{eq:Cherenkov_condition}) with respect
to the arc length $s$ results in \begin{equation}
\left(\bi{v}_{{\rm {g}}}(\bi{k}(s))-\bi{v}\right)\cdot\frac{\partial\bi{k}}{\partial s}=0.\label{eq:stat_add}\end{equation}
 Combining the stationary phase condition (\ref{eq:stat_4}) with
relation (\ref{eq:stat_add}) and taking into account the definition
of ${\rm BZ}_{1}$ (\ref{eq:stationary_3_ie}), one can see that for
the stationary Bloch modes the vector $\bi{v}_{{\rm {g}}}(\bi{k}^{\nu})-\bi{v}$
must be parallel to the vector $\bi{r}-\bi{v}t$. In other words,
only the eigenmodes whose group velocities in the coordinate frame
moving with the point (line) charge pointing towards an observation
direction in this coordinate frame, contribute to the Cherenkov radiation
in the far-field zone. This statement is illustrated in figure~\ref{fig:group_velocity}.
The group velocity contour corresponding to the integration contour
$\mathcal{C}_{m}$ is shown. The color coding is used as in Fig.~\ref{fig:band_structure},
the same color corresponds to the same frequency and consequently
to the same wave vector. Main contributions to the integral (\ref{eq:I_minus_para_1})
for two different observation directions are depicted. For the direction
$\theta_{1}$ only one Bloch mode with group velocity $\bi v_{{\rm g}}^{\nu_{1}}$
satisfies the stationary phase condition, while for the direction
$\theta_{2/3}$ there are two modes with group velocities $\bi v_{{\rm g}}^{\nu_{2}}$
and $\bi v_{{\rm g}}^{\nu_{3}}$ fulfilling the condition. There does
not exist Bloch modes satisfying the stationary phase condition for
forward observation directions characterized by angles smaller than
$\theta_{1}$.

The integral (\ref{eq:I_minus_para_1}) can be further approximated
by expanding the phase $h\left(s\right)$ near the stationary points
$\bi{k}(s^{\nu})=\bi{k}^{\nu}$ in a Taylor series up to quadratic
order\begin{equation}
h\left(s\right)\approx\bi{k}^{\nu}\cdot\bi{r}-\omega_{n}(\bi{k}^{\nu})t+\frac{1}{2}\left(\bi{r}-\bi{v}t\right)\cdot\frac{\partial^{2}\bi{k}^{\nu}}{\partial s^{2}}\left(s-s^{\nu}\right)^{2}\label{eq:taylor_1D}\end{equation}
 and extending the integration range to the whole real axis \begin{eqnarray}
I_{-} & \approx & 2\pi\sum_{\nu}\frac{\bi{e}_{n}\left(\bi{k}^{\nu},\bi{r}\right)c_{nm}(\bi{k}^{\nu})}{\left|\bi{v}_{{\rm {g}}}(\bi{k}^{\nu})-\bi{v}\right|}\exp\left\{ \rmi\left[\bi{k}^{\nu}\cdot\bi{r}-\omega_{n}(\bi{k}^{\nu})t\right]\right\} \nonumber \\
 &  & \cdot\int_{-\infty}^{\infty}\rmd s\,\exp\left\{ \frac{\rmi}{2}\left[\left(\bi{r}-\bi{v}t\right)\cdot\frac{\partial^{2}\bi{k}^{\nu}}{\partial s^{2}}\right]\left(s-s^{\nu}\right)^{2}\right\} .\label{eq:I_minus_approx}\end{eqnarray}
 Where the summation is taken over all stationary points $\nu$. The
resulting integral can be evaluated analytically \cite{gradshteyn07}
\begin{equation}
\int_{-\infty}^{\infty}dt\,\rme^{\rmi bt^{2}}=\sqrt{\frac{\pi}{|b|}}\exp\left[\rmi\frac{\pi}{4}\mathrm{sign}(b)\right],\label{eq:quadratic_integral}\end{equation}
 which together with the relation \begin{equation}
\left(\bi{r}-\bi{v}t\right)\cdot\frac{\partial^{2}\bi{k}^{\nu}}{\partial s^{2}}={\rm sign}\left(\left(\bi{r}-\bi{v}t\right)\cdot\frac{\partial^{2}\bi{k}^{\nu}}{\partial s^{2}}\right)\left|\left(\bi{r}-\bi{v}t\right)\right|\left|\frac{\partial^{2}\bi{k}^{\nu}}{\partial s^{2}}\right|\label{eq:curvature_relation}\end{equation}
 results in the final expression for the Cherenkov electric field
in far-field zone for a 2D periodic dielectric medium\begin{eqnarray}
\bi{E}^{(2D)}(\bi{r},t) & \approx & -\frac{q\left|\bi{v}\right|}{\varepsilon_{0}\sqrt{2\pi\left|\bi{r}-\bi{v}t\right|}}{\rm Re}\Biggl\{\sum_{n,\nu}\frac{\bi{e}_{n}\left(\bi{k}^{\nu},\bi{r}\right)c_{nm}(\bi{k}^{\nu})}{\left|\bi{v}_{{\rm {g}}}(\bi{k}^{\nu})-\bi{v}\right|\sqrt{\mathcal{K}^{\nu}}}\nonumber \\
 &  & \cdot\exp\left\{ \rmi\left[\bi{k}^{\nu}\cdot\bi{r}-\omega_{n}(\bi{k}^{\nu})t\right]\right\} \nonumber \\
 &  & \cdot\exp\left\{ \rmi\left[\frac{\pi}{4}{\rm sign}\left(\left(\bi{r}-\bi{v}t\right)\cdot\vec{\mathcal{K}}^{\nu}\right)\right]\right\} \Biggr\}.\label{eq:endergebnis_2d}\end{eqnarray}
 Here $\mathcal{K}^{\nu}=\left|\vec{\mathcal{K}}^{\nu}\right|=\left|\frac{\partial^{2}\bi{k}^{\nu}}{\partial s^{2}}\right|$
is the curvature of the contour $\mathcal{C}_{m}$ at the stationary
point $\bi{k}^{\nu}$. Relation (\ref{eq:curvature_relation}) holds
because of $\frac{\partial\bi{k}}{\partial s}\perp\frac{\partial^{2}\bi{k}}{\partial s^{2}}$
and (\ref{eq:stat_4}).

\subsection{3D periodic media}

By introducing a 2D coordinate system with unit vectors $\frac{\partial\bi{k}}{\partial s_{1}}$
and $\frac{\partial\bi{k}}{\partial s_{2}}$ tangential to the integration
surface $\mathcal{C}_{m}$, the surface integral (\ref{eq:I_minus_final})
with $d=2$ can be expressed as\begin{eqnarray}
I_{-} & \approx & 2\pi\sum_{m}\int\int\rmd s_{1}\rmd s_{2}\left|\frac{\partial\bi k}{\partial s_{1}}\times\frac{\partial\bi{k}}{\partial s_{2}}\right|\frac{\bi{e}_{n}\left(\bi{k}(\bi{s}),\bi{r}\right)c_{nm}(\bi{k}(\bi{s}))}{\left|\bi{v}_{{\rm {g}}}(\bi{k}(\bi{s}))-\bi{v}\right|}\nonumber \\
 &  & \quad\cdot\exp\left[\rmi\left(\bi{k}(\bi{s})\cdot\bi{r}-\omega_{n}(\bi{k}(\bi{s}))t\right)\right].\label{eq:I_minus_param_3D}\end{eqnarray}
 where $\bi{s}\in\mathbb{R}^{2}$. Similar to the 2D case, the main
contribution to the integral comes from the neighborhood of \emph{k}-points
$\bi{k}(\bi{s}^{\nu})=\bi{k}^{\nu}$ where the phase $h\left(\bi{s}\right)=\bi{k}(\bi{s})\cdot\bi{r}-\omega_{n}(\bi{k}(\bi{s}))t$
is stationary\begin{equation}
\left(\bi{r}-\bi{v}t\right)\cdot\frac{\partial\bi{k}^{\nu}}{\partial s_{i}}=0,\label{eq:stat_3D}\end{equation}
 with $i\in\left\{ 1,2\right\} $. Choosing a local coordinate system
at the stationary point $\bi{k}^{\nu}$ with basis vectors $\frac{\partial\bi{k}}{\partial\xi_{1}}$
and $\frac{\partial\bi{k}}{\partial\xi_{2}}$ along the main directions
of the surface curvatures the following form of the Taylor expansion
of the phase $h\left(\bi{s}\right)$ can be used to evaluate the integral
(\ref{eq:I_minus_param_3D})\begin{eqnarray}
h\left(\xi_{1},\xi_{2}\right) & \approx & \bi{k}^{\nu}\cdot\bi{r}-\omega_{n}(\bi{k}^{\nu})t\nonumber \\
 & + & \frac{1}{2}\left(\bi{r}-\bi{v}t\right)\cdot\left(\frac{\partial^{2}\bi{k}^{\nu}}{\partial\xi_{1}^{2}}\left(\xi_{1}-s_{1}^{\nu}\right)^{2}+\frac{\partial^{2}\bi{k}^{\nu}}{\partial\xi_{2}^{2}}\left(\xi_{2}-s_{2}^{\nu}\right)^{2}\right),\label{eq:taylor_3D}\end{eqnarray}
 where $\frac{\partial^{2}\bi{k}^{\nu}}{\partial\xi_{1}^{2}}$ and
$\frac{\partial^{2}\bi{k}^{\nu}}{\partial\xi_{2}^{2}}$ are the main
curvatures $\mathcal{K}_{1}^{\nu}$ and $\mathcal{K}_{2}^{\nu}$ of
the integration surface $\mathcal{C}_{m}$. The integration limit
is than extended to the whole real plane. Using twice the relation
(\ref{eq:quadratic_integral}) the Cherenkov electric field in far-field
zone for a 3D periodic dielectric medium reads\begin{eqnarray}
\bi{E}^{(3D)}(\bi{r},t) & \approx & -\frac{q\left|\bi{v}\right|}{2\pi\varepsilon_{0}\left|\bi{r}-\bi{v}t\right|}{\rm Re}\Biggl\{\sum_{n,\nu}\frac{\bi{e}_{n}\left(\bi{k}^{\nu},\bi{r}\right)c_{nm}(\bi{k}^{\nu})}{\left|\bi{v}_{{\rm {g}}}(\bi{k}^{\nu})-\bi{v}\right|\sqrt{\mathcal{K}_{1}^{\nu}\mathcal{K}_{2}^{\nu}}}\nonumber \\
 &  & \cdot\exp\left\{ \rmi\left[\bi{k}^{\nu}\cdot\bi{r}-\omega_{n}(\bi{k}^{\nu})t\right]\right\} \nonumber \\
 &  & \cdot\exp\left\{ \rmi\left[\frac{\pi}{4}{\rm sign}\left(\left(\bi{r}-\bi{v}t\right)\cdot\vec{\mathcal{K}}_{1}^{\nu}\right)\right]\right\} \nonumber \\
 &  & \cdot\exp\left\{ \rmi\left[\frac{\pi}{4}{\rm sign}\left(\left(\bi{r}-\bi{v}t\right)\cdot\vec{\mathcal{K}}_{2}^{\nu}\right)\right]\right\} \Biggr\},\label{eq:endergebnis_3d}\end{eqnarray}
 with the product of the main curvatures $\mathcal{K}_{1}^{\nu}\mathcal{K}_{2}^{\nu}$
being the Gaussian curvature of the surface $\mathcal{C}_{m}$ defined
by the generalized Cherenkov condition (\ref{eq:Cherenkov_condition}).
The vectors $\vec{\mathcal{K}}_{i}$ are defined by $\vec{\mathcal{K}}_{i}=\frac{\partial^{2}\bi{k}^{\nu}}{\partial\xi_{i}^{2}}$.

\section{Discussion and numerical examples\label{sec:Numerical-example}}

Formulas (\ref{eq:E_final}), (\ref{eq:endergebnis_2d}) and (\ref{eq:endergebnis_3d})
constitute the main result of the present work. In the far-field zone
the electric field generated by a point (line) charge uniformly moving
in a 3D (2D) periodic medium is dominated by a small number of Bloch
eigenmodes of the medium. To calculate the far-field, these Bloch
modes (their wave vectors) should be calculated as a solution of the
generalized Cherenkov condition (\ref{eq:Cherenkov_condition}). In
turn, the spatial variation of the Cherenkov radiation is dominated
(i) by the interference of these Bloch modes at the observation point
and (ii) by the topology of the dispersion relation at the Cherenkov
surface (contour) in Eqs. (\ref{eq:E_final}) and (\ref{eq:endergebnis_3d})
(Eqs. (\ref{eq:E_final}) and (\ref{eq:endergebnis_2d})).

In what follows, we apply formulas (\ref{eq:E_final}) and (\ref{eq:endergebnis_2d})
to study the spatial distribution of Cherenkov radiation in the 2D
photonic crystal depicted in Fig.~\ref{fig:sketch}. The line charge
oriented perpendicular to the periodicity plane of the crystal moves
along the $x$-axis with a velocity $v=0.15c$, staying always in
the space between air holes. The corresponding current density (\ref{eq:current_time})
couples only to Bloch eigenmodes with an electric field polarized
in the periodicity plane (TE polarization). The first TE photonic
band of the considered crystal is presented in figure~\ref{fig:band_structure}.
The band structure as well as group velocities and Bloch eigenmodes
were calculated using the plane wave expansion method \cite{johnson01}.

In order to calculate both the electric field (\ref{eq:E_final})
and its approximation (\ref{eq:endergebnis_2d}) the set of wave vectors
contributing to the far-field should be calculated. This set can be
found as a numerical solution of the generalized Cherenkov condition
(dashed line in Fig.~\ref{fig:band_structure}). In contrast to the
homogeneous medium case, such a solution does exist for an arbitrary
charge velocity \cite{kremers09}.

Having the set of wave vectors $\mathcal{C}_{m}$, the Bloch modes
$\bi e_{\bi kn}$ and the Fourier coefficient $c_{nm}(\bi k)$ can
be calculated. The Fourier coefficients $c_{nm}(\bi k)$ give the
coupling strength between the current associated with the moving charge
and the Bloch eigenmodes. As it can be seen from the definition (\ref{eq:def_of_c_m}),
with increasing index $|m|$ the Fourier coefficients become smaller,
reducing the contribution of the higher frequencies to the Cherenkov
radiation in the far-field zone. Finally, the Cherenkov field (\ref{eq:E_final})
can be calculated by direct numerical integration.

\begin{figure}
\begin{centering}
\includegraphics[width=0.5\textwidth]{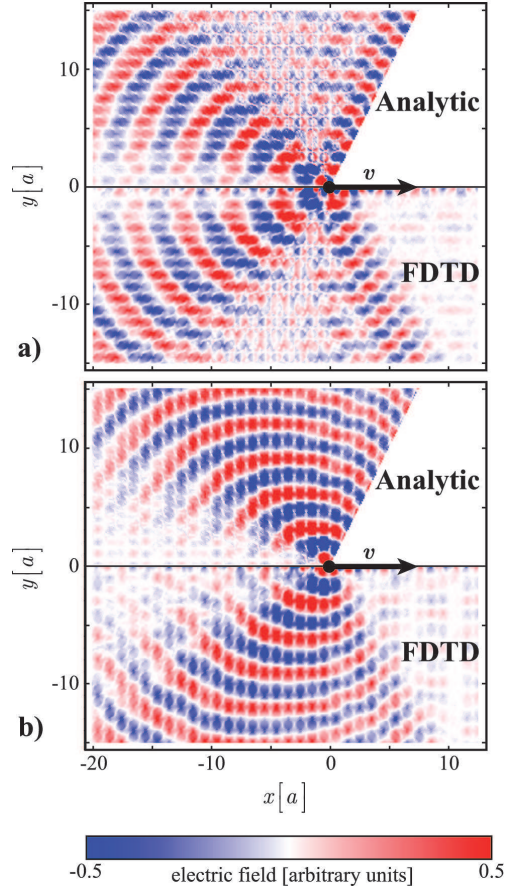} 
\par\end{centering}

\caption{$E_{x}(\bi r,t=0)$ (top panel) and $E_{y}(\bi r,t=0)$ (bottom panel)
components of the Cherenkov electric field for a charge velocity $v=0.15c$.
The results of the stationary phase approximation (\ref{eq:endergebnis_2d})
and direct numerical integration using FDTD methods are shown.\label{fig:numerics_2D}}

\end{figure}

To calculate the stationary phase approximation of the far-field (\ref{eq:endergebnis_2d}),
stationary wave vectors $\bi{k}^{\nu}$ should be calculated for a
given observation direction $\bi{r}-\bi{v}t$. This can be done by
parameterizing the Cherenkov contour by the arc length $s$ and looking
for all Bloch modes whose group velocities (Fig.~\ref{fig:group_velocity})
satisfy the stationary phase condition $\left(\bi{v}_{{\rm {g}}}(\bi{k}^{\nu})-\bi{v}\right)\upuparrows\left(\bi{r}-\bi{v}t\right)$.
Furthermore, the Bloch modes, the Fourier coefficient and the curvature
of the Cherenkov contour (see Appendix) should be calculated and summed
for these stationary points only. This reduce the computational demands
considerably.

In figure~\ref{fig:numerics_2D}, a numerical calculations of the
stationary phase approximation of the far-field is presented for $v=0.15c$.
The $E_{x}(\bi r,t=0)$ and $E_{y}(\bi r,t=0)$ components of the
electric field are shown in the upper halves of the top and bottom
panels, respectively. Only contributions from the first three sections
of the Cherenkov contour corresponding to $m=0,\,-1,\,-2$ have been
analyzed. In the case of a homogeneous medium, the standard Cherenkov
condition would impose a minimal charge velocity above which Cherenkov
radiation is possible $v_{\mathrm{min}}\geq c/n_{\mathrm{eff}}\approx0.457c$.
Here an effective refractive index of the considered periodic medium
is asymptotically equal to $n_{\mathrm{eff}}=\sqrt{\varepsilon_{\mathrm{eff}}}\thickapprox2.186$.
Although the velocity of the line charge is considerably smaller than
$v_{\mathrm{min}}$, in the periodic medium the non-evanescent field
can be clearly seen far apart from the charge trajectory (Fig.~\ref{fig:numerics_2D}).
Another characteristic feature of the spatial distribution of the
Cherenkov radiation for $v=0.15c$ is a backward-pointing radiation
cone \cite{Luo03}. The field in the forward direction for observation
angles smaller than $\theta_{1}$ is zero.

\begin{figure}
\begin{centering}
\includegraphics[width=0.9\textwidth]{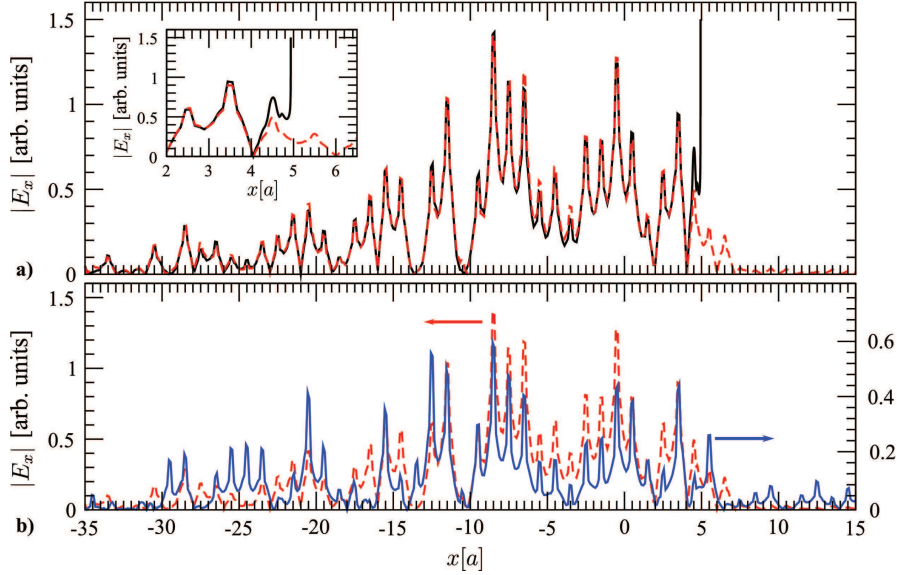} 
\par\end{centering}

\caption{Comparison of the integral representation of the far-field (\ref{eq:E_final})
(dashed red line) with the stationary phase approximation (\ref{eq:endergebnis_2d})
(black solid line) and FDTD calculations (blue solid line). An absolute
value of the horizontal component of the electric field is shown 10
lattice constants apart from the charge trajectory.\label{fig:numerics_cut}}

\end{figure}

The zero field within the backward-pointing radiation cone is associated
with the absence of stationary solutions for observation directions
in the cone (Fig.~\ref{fig:group_velocity}). This is a direct consequence
of the stationary phase approximation. In figure~\ref{fig:numerics_cut}
a), the stationary phase approximation of the far-field 10 lattice
constants apart from the trajectory is compared with the direct numerical
integration of the integral representation of the field (\ref{eq:E_final}).
An excellent agreement between these two solutions can be seen up
to the Cherenkov cone $\theta_{1}$, where the stationary phase approximation
breaks down. The cone angle corresponds to the fold in the group velocity
contour (Fig.\ \ref{fig:group_velocity}). The curvature of the Cherenkov
contour is zero at the corresponding wave vector and the Taylor expansion
(\ref{eq:taylor_1D}) fails to reproduce the contour accurately in
the vicinity of this point. Zero curvature leads to the diverging
field (Fig.~\ref{fig:numerics_cut}-inset). The field calculated
with Eq. (\ref{eq:E_final}) is small but finite in forward directions.

To substantiate our analytical calculations, a direct numerical integration
of Maxwell's equations (\ref{eq:maxwell_rotE}-\ref{eq:maxwell_rotH})
using rigorous finite-differences time-domain (FDTD) method \cite{taflove95}
has been performed. The corresponding FDTD field distributions are
presented in the lower halves of the top and bottom panels of figure~\ref{fig:numerics_2D}.
$200a\,\times70a$ lattice of holes with discretization $\Delta=a/18$
has been used for the FDTD calculation. The simulation domain is surrounded
by a $3a$-wide perfectly matched layer (PML). The integration time
step is set to $98\%$ of the Courant value. In order to describe
a continuous movement of the charge, the source (\ref{eq:current_time})
is modeled by a discrete Gaussian in the $x$-direction with a standard
deviation $\sigma=\Delta$ and by an appropriately normalized Kronecker
delta in the $y$-direction\[
j(x_{i}\Delta,y_{j}\Delta)=\frac{\delta_{jk}}{\sqrt{2\pi}\Delta^{2}}\exp\left[-\frac{\left(x_{i}\Delta-vt\right)^{2}}{2\Delta^{2}}\right]\]
 where $y_{k}=35a$ is the center of the crystal in the vertical direction.
In order to compare directly FDTD results with the predictions of
formulas~(\ref{eq:E_final}) and (\ref{eq:endergebnis_2d}), the
static contribution as well as the higher frequency contributions
to the FDTD field have been filtered out. An overall good agreement
between the results of the analytical and direct numerical calculation
is obtained (Fig.~\ref{fig:numerics_2D} and \ref{fig:numerics_cut}).
The FDTD calculation follows nicely the main characteristic of the
analytically predicted field. The difference in the absolute values
between FDTD and analytical calculations (Fig.~\ref{fig:numerics_cut})
can be associated with the residual reflections from the perfectly
matched layer.

\section{Conclusion\label{sec:Conclusions}}

In conclusion, we have developed an analytical theory of the Cherenkov
radiation in the far-field zone. The field far apart from the charge
trajectory can be calculated as a surface (contour) integral over
a small fraction of the first Brillouin Zone in a 3D (2D) dielectric
medium. We have shown that the main contribution to this integral
comes from a small and discrete number of $k$-points. This opens
the possibility to calculate the integral approximately, but with
high accuracy. We have also shown, that the spatial variation of the
Cherenkov radiation in the far-field is due to the interference of
a few Bloch eigenmodes as well as the topological properties of the
Cherenkov surface (contour). This has been defined as a manifold of
all $k$-points contributing to the Cherenkov radiation for a given
charge velocity. Simple formulas have been derived for the Cherenkov
far-field both in 3D and 2D cases. We have compared the developed
analytical theory with numerically rigorous FDTD calculations. A good
agreement between these two methods has been demonstrated.

\ack{We are grateful to Sergei Zhukovsky and Fuh Chuo Evaristus
for fruitful discussions. Financial support from the Deutsche Forschungsgemeinschaft
(DFG FOR 557) is acknowledged.}

\appendix
\setcounter{section}{1}

\section*{Appendix}

To calculate the curvature using its definition \[
\mathcal{K}(s)=\left|\frac{\partial^{2}\bi{k}(s)}{\partial s^{2}}\right|\]
 one should numerically evaluate the second derivative of the wave
vector $\bi{k}(s)$ on the Cherenkov contour. This is typically associated
with a large numerical error. In what follows we propose to use an
alternative method to calculate the curvature which usually results
in smaller numerical error and involves calculations of the first
derivative of the wave vector on the contour and the second derivatives
of the dispersion relation $\omega_{n}(\bi k)$.

Taking the derivative of (\ref{eq:stat_add}) with respect to $s$
we obtain\begin{equation}
\frac{\partial^{2}\bi{k}}{\partial s^{2}}\cdot\left(\bi{v}_{{\rm {g}}}(\bi{k}(s))-\bi{v}\right)+\frac{\partial\bi{k}}{\partial s}\cdot\frac{\partial}{\partial s}\bi{v}_{{\rm {g}}}(\bi{k}(s))=0\label{eq:appd1}\end{equation}
 The second term in (\ref{eq:appd1}) can be rewritten as\begin{eqnarray*}
\frac{\partial\bi{k}}{\partial s}\cdot\frac{\partial}{\partial s}\bi{v}_{{\rm {g}}}(\bi{k}(s)) & = & \sum_{i}\frac{\partial k_{i}}{\partial s}\left[\frac{\partial}{\partial s}\frac{\partial}{\partial k_{i}}\omega_{n}(\bi{k})\right]\\
 & = & \sum_{i,j}\frac{\partial k_{i}}{\partial s}\frac{\partial^{2}}{\partial k_{j}\partial k_{i}}\omega_{n}(\bi{k})\frac{\partial k_{j}}{\partial s}\\
 & = & \left(\frac{\partial\bi{k}}{\partial s}\right)^{T}\mathcal{H}_{\omega_{n}}(\bi{k})\frac{\partial\bi{k}}{\partial s}=\mathcal{M},\end{eqnarray*}
 where $\mathcal{H}_{\omega_{n}}(\bi{k})$ $ $is a Hessian matrix
of the dispersion relation $\omega_{n}(\bi k)$. We can now rewrite
(\ref{eq:appd1}) in the form\[
\frac{\partial^{2}\bi{k}}{\partial s^{2}}\cdot\left(\bi{v}_{{\rm {g}}}(\bi{k}(s))-\bi{v}\right)=\mathcal{M}\]
 or\[
\mathcal{K}\left|\bi{v}_{{\rm {g}}}(\bi{k}(s))-\bi{v}\right|{\rm sign}\left[\vec{\mathcal{K}}\cdot\left(\bi{v}_{{\rm {g}}}(\bi{k}(s))-\bi{v}\right)\right]=\mathcal{M}\]
 with $\vec{\mathcal{K}}=\frac{\partial^{2}\bi{k}}{\partial s^{2}}.$
Finally the following expression for the curvature is obtained\begin{equation}
\mathcal{K}=\frac{\left|\mathcal{M}\right|}{\left|\bi{v}_{{\rm {g}}}(\bi{k}(s))-\bi{v}\right|}.\label{eq:appd2}\end{equation}

\end{document}